\documentclass[aps,prl,twocolumn,floatfix,floats,showpacs,superscriptaddress,raggedbottom]{revtex4-1}
\usepackage{graphicx,latexsym}
\usepackage{dcolumn}
\usepackage{amsmath}
\usepackage{amssymb,bm}
\usepackage{times}
\usepackage{color}
\usepackage[normalem]{ulem}

\def\fig#1{Fig.\ \ref{#1}}

\usepackage{hyperref}
\hypersetup{
    pdfnewwindow=true,       
    colorlinks=true,         
    linkcolor=blue,          
    citecolor=blue,          
    filecolor=magenta,       
    urlcolor=black           
}

\begin{document}

\title{Cavity-photon controlled thermoelectric transport through a quantum wire}

\author{Nzar Rauf Abdullah}
\email{nzar.r.abdullah@gmail.com}
\affiliation{Physics Department, Faculty of Science and Science Education, School of Science, 
             University of Sulaimani, Kurdistan Region, Iraq}
\affiliation{Science Institute, University of Iceland, 
             Dunhaga 3, IS-107 Reykjavik, Iceland}

\author{Chi-Shung Tang}
\affiliation{Department of Mechanical Engineering,
  National United University, 1, Lienda, Miaoli 36003, Taiwan}

\author{Andrei Manolescu}
\affiliation{Reykjavik University, School of Science and Engineering,
              Menntavegur 1, IS-101 Reykjavik, Iceland}

\author{Vidar Gudmundsson}
\email{vidar@hi.is}
 \affiliation{Science Institute, University of Iceland,
        Dunhaga 3, IS-107 Reykjavik, Iceland}

%

\begin{abstract}

We investigate the influence of a quantized photon field on thermoelectric transport 
of electrons through a quantum wire embedded in a photon cavity. 
The quantum wire is connected to two electron reservoirs at different temperatures 
leading to a generation of a thermoelectric current.
The transient thermoelectric current strongly depends on the photon energy and the number of photons initially 
in the cavity. Two different regimes are studied, off-resonant and resonant polarized fields,
with photon energy smaller than, or equal to the energy spacing between some of the lowest states 
in the quantum wire.
We observe that the current is inverted for the off-resonant photon field due to participation 
of photon replica states in the transport.
A reduction in the current is recorded for the resonant photon field, a direct consequence of the 
Rabi-splitting. 

\end{abstract}



\maketitle

%
%

\section{Introduction}\label{Sec:Introduction}
Thermoelectric transport is a subject of intense study  
for future energy harvesting devices~\cite{Shakouri41.399,Majumdar303.777}.
Low-dimensional electronic systems have a potential to improve 
thermoelectric efficiency compared to bulk electronic materials 
due to their highly peaked density of states for system sizes  
in the range of nanometers~\cite{Heremans25072008}.
A thermoelectric current (TEC) can be generated by a temperature gradient $\Delta T$ in an electronic systems.
In the linear response regime, the temperature gradient approaches zero
and the thermoelectric efficiency is characterized by the dimensionless figure of merit $ZT$ \cite{PhysRevB.47.12727}.
In the nonlinear response regime, the thermoelectric efficiency is represented by 
the bias voltage $\Delta V$ generated by $\Delta T$~\cite{PhysRevLett.99.027203}.
In both regimes, the thermal efficiency can be high in nanodevices.

Since the 1990's, the thermoelectric transport has been investigated in several quantum systems, such as 
single quantum dots~\cite{PhysRevB.46.9667,PhysRevLett.86.280,1367-2630-14-3-033041}, 
double quantum dots~\cite{PhysRevB.80.195409}, 
and quantum wires~\cite{PhysRevB.64.045324}.
The Coulomb interaction between charge carriers influences the thermal transport through
a quantum dot and forms plateaus in the TEC~\cite{PhysicaE.53.178}.
The thermoelectric effect through a serial double quantum dot weakly coupled to ferromagnetic
leads has been investigated, and the influence of temperature and inter-dot tunneling on the 
figure of merit has been demonstrated~\cite{Tagani2012914}.

Another interesting aspect of this issue is the importance of photon radiation to control
thermal transport. Recently, time-dependent photon radiation has been used to enhance 
the heat and thermoelectric transport~\cite{PhysRevB.87.085427, Hochbaum455.778}. 
Photonic heat current through an arbitrary circuit element coupled to two dissipative reservoirs 
has been explored at finite temperatures~\cite{PhysRevLett.100.155902}. Transfer of heat via photons between 
two metals is reported to study photonic power where the metals are coupled 
by a circuit containing linear reactive impedance~\cite{PhysRevB.83.125113}.

Here, we describe the TEC in a quantum wire coupled to two electron reservoirs with
different temperatures and the same chemical potential. The temperature gradient causes electron flow from 
the leads to the quantum wire and vice versa. 
In addition, the quantum wire is coupled to a cavity field with photons polarized in the  
direction of electron propagation in the quantum wire. A generalized master equation is used to calculate 
the time-dependent evolution of electrons in the quantum wire~\cite{Nzar.25.465302, nzar27.015301}.
We show how TEC can be controlled by cavity parameters such as the photon energy 
in off-resonant and on-resonant photon fields.


\section{Method and Theory}\label{Sec:Results}
We assume a short quantum wire is formed in a two-dimensional electron gas in the $x$- and $y$-plane.
The quantum wire is hard-wall confined at the ends in the $x$-direction and parabolically confined in the $y$-direction.
The wire with length $L_x = 150$~nm is weakly coupled to two leads held at different temperatures.
The total system is exposed to an external perpendicular static magnetic field $B = 0.1$~T
with cyclotron energy $\hbar\omega_c = 0.172$~meV.
The transverse confinement energy of the electrons in the quantum wire is equal to that of the 
leads $\hbar \Omega_0 = \hbar \Omega_l = 2.0$~meV,
where $l$ refers to the left (L), or the right (R) lead. The effective confinement frequency is 
$\Omega_w = \sqrt{\Omega_0^2 + \omega_c^2}$.
The photons in the cavity are linearly polarized in the $x$-direction. 
We tune the photon energy, and the number of photons initially in the cavity
to control the transient TEC in the quantum wire.

\begin{figure}[htbq]
  \includegraphics[width=0.4\textwidth]{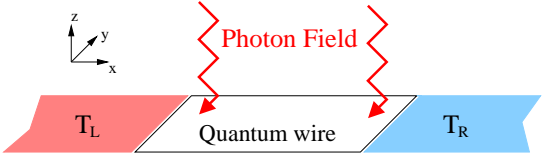}
 \caption{(Color online) Schematic diagram of a quantum wire (white color) connected to a left lead (pink color) with 
                         temperature $T_L$, and a right lead (blue color) with temperature $T_R$. The photon field 
                         is represented by the red zigzag arrows.}
\label{fig01}
\end{figure}

\fig{fig01} shows a schematic diagram of the quantum wire (white color) coupled to the left lead (red color) 
with temperature $T_L$, 
and the right lead (blue color) with temperature $T_R$. The red zigzag arrows indicate the polarized photon field. 

The time-evolution of the many-body density operator of the system is governed by the Liouville-von Neumann 
equation. Due to its complexity a generalized master equation for the reduced density operator is derived by 
projecting the system description onto the central system, the short quantum wire \cite{Nakajima20.948,Zwanzing.33.1338}.
As we are dealing with off- and on-resonance processes we include both the para- and the diamagnetic 
electron-photon interactions without the rotating wave approximation \cite{PhysRevE.86.046701}. 
The electron-electron and the electron-photon interactions are treated by exact diagonalization 
in appropriately truncated Fock-spaces. Our interest in
the transient behavior of the system requires a non-Markovian approach valid to a weak coupling
of the leads to the central system \cite{Vidar61.305}.  

\subsection{Off-resonance photon field}
In this section, we assume the photon energy is smaller than the energy spacing between the three lowest energy states
of the quantum wire. For instance, the photon energy is $\hbar\omega_{\gamma} < E_{1}-E_{0}$
where $E_{0}$($E_{1}$) is the energy of the ground state (first-excited state) of the quantum wire, respectively.

\begin{figure}[htbq]
  \includegraphics[width=0.2\textwidth,angle=0,bb=60 70 200 290]{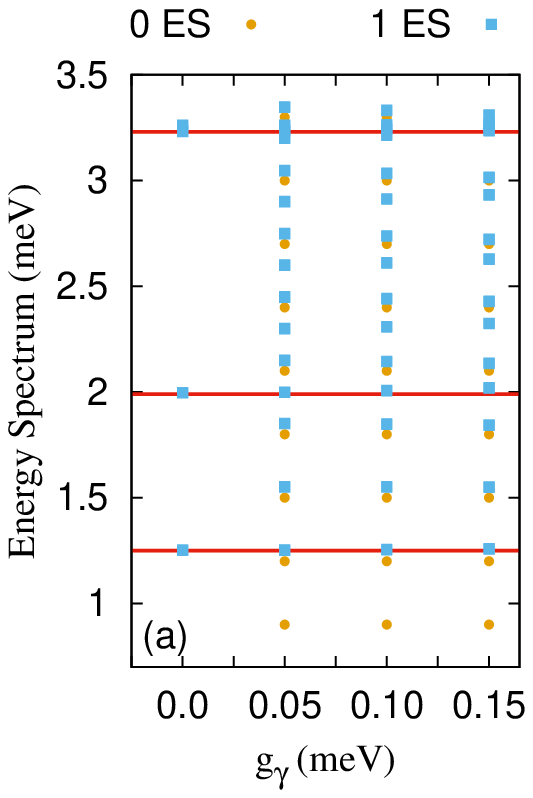}
  \includegraphics[width=0.2\textwidth,angle=0,bb=90 68 240 295]{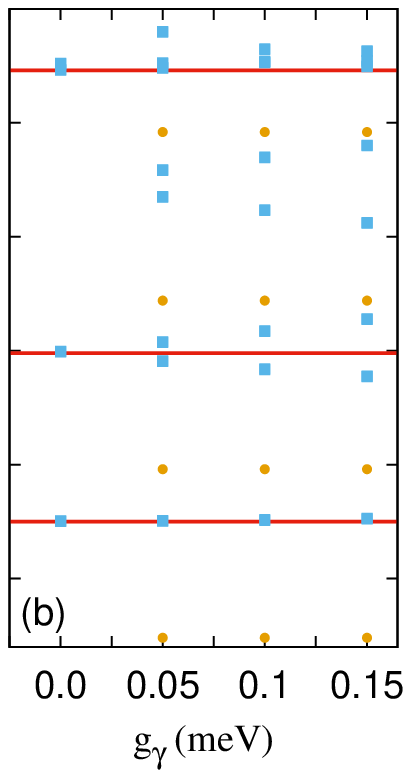}
 \caption{(Color online) Energy spectrum of the quantum wire versus electron-photon coupling strength $g_{\gamma}$.
         0ES (golden circles) are zero-electron states, 1ES (blue squares) are one-electron states 
         and the horizontal lines (red lines) display the location of the resonances 
         of the leads with the three lowest one-electron states in the case of off-resonance (a) and resonance (b) photon field.
         The photon energy is $\hbar\omega_{\gamma} = 0.3$~meV, $N_{\gamma} = 2$, and the photons are linearly polarized 
         in the $x$-direction. The magnetic field is $B = 0.1~{\rm T}$, and $\hbar \Omega_0 = 2.0~{\rm meV}$.}
\label{fig02}
\end{figure}

Figure~\ref{fig02}(a) shows the energy spectrum versus electron-photon coupling strength $g_{\gamma}$, 
where 0ES (golden circles) are zero-electron states, 1ES (blue squares) are one-electron states and the 
horizontal lines (red lines) indicate the location of the resonance condition for the 
lowest three one-electron states with the chemical potentials of the leads, $\mu_L=\mu_R=E_\mu$, at
$g_{\gamma}=0$.
We start with no electron-photon coupling, $g_{\gamma} = 0.0$~meV. 
In this case, we concentrate our attention on the three lowest states in a selected range of the energy spectrum:
the ground state, the first-excited state, and the second-excited state with energy values $E_0 = 1.25$~meV, $E_1 = 1.99$~meV, 
and $E_2 = 3.23$~meV, respectively. In the presence of electron-photon coupling with off-resonant photon field, photon replicas 
of the above mentioned three states are formed.
The separation between the photon replica states is approximately equal to the 
photon energy for the selected values of the electron-photon coupling strength used here. 
The photon replicas play an important role in the thermoelectric transport.

\begin{figure}[htbq]
  \includegraphics[width=0.5\textwidth,angle=0,bb=50 95 410 200]{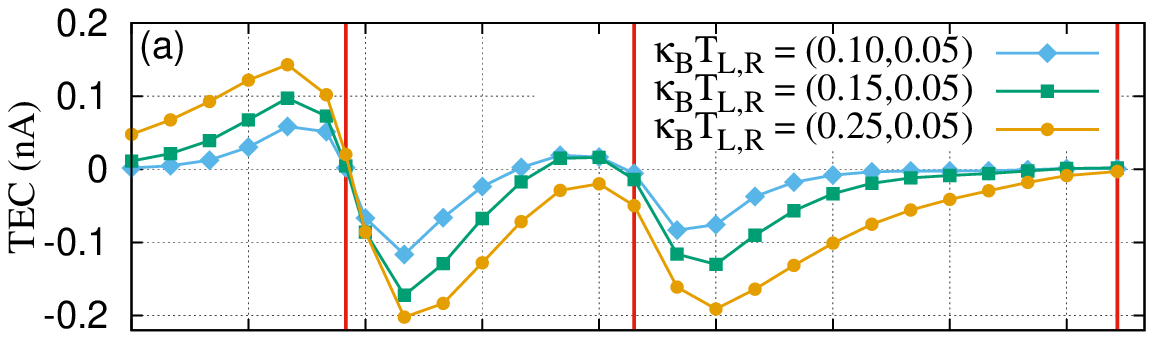}\\
    \includegraphics[width=0.5\textwidth]{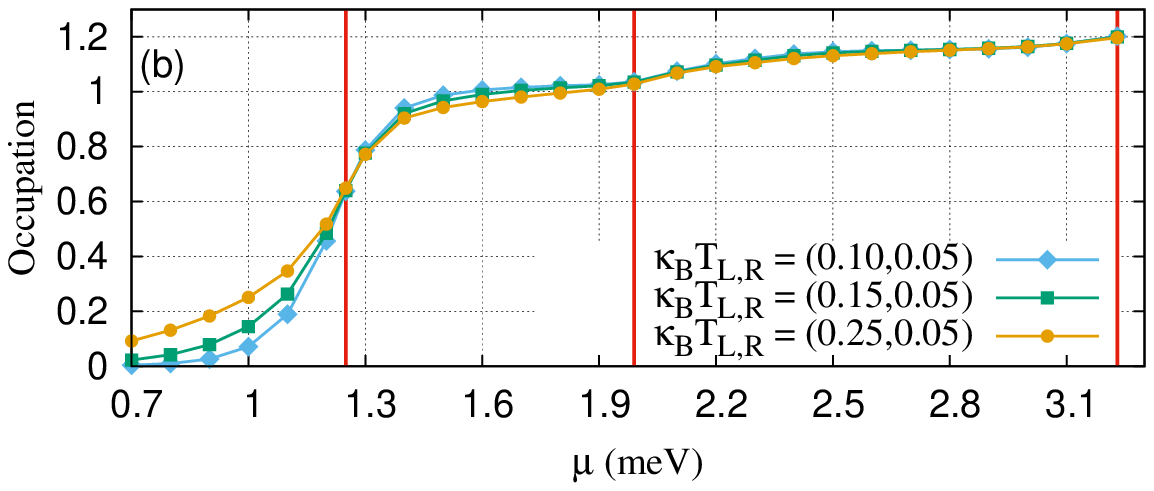}
 \caption{(Color online) TEC (a), and 
               occupation (b) as functions of the chemical potential $\mu = \mu_L = \mu_R$ plotted at time $t = 220$~ps.
               The temperature of the right lead is fixed at $T_R = 0.58$~K implying thermal energy $0.05$~meV,
               and varying the temperature of the left lead to $T_L = 1.16$, $1.742$ and 
               $2.901$~K implying thermal energies $0.1$ (blue diamonds), $0.15$ (green squares),
               and $0.25$~meV (golden circles), respectively.
               The red vertical lines show the resonance condition for the ground state at $\mu = 1.25$~meV, 
               the first-excited state at $\mu = 1.99$~meV,
               and the second-excited state at $\mu = 3.23$~meV, respectively.
               The magnetic field is $B = 0.1~{\rm T}$, and
               $\hbar \Omega_0 = 2.0~{\rm meV}$.}
\label{fig03}
\end{figure}

Figure \ref{fig03} demonstrates the TEC (a) and the occupation (b) 
for the three lowest states of the quantum wire in the case of no electron-photon coupling, $g_{\gamma} = 0.0$~meV.
We fix the temperature of the right lead at $k_B T_R = 0.05$~meV, 
and vary the temperature of the left lead to $k_B T_L = 0.1$ (blue diamonds), $0.15$ (green squares) and $0.25$~meV (golden circles).
In \fig{fig03}(a) the TEC as a function of chemical potential $\mu = \mu_L = \mu_R$ is plotted 
at time $t = 220$~ps. At this time point the system is in the late transient regime very close to a 
steady state.
The concept of a TEC can be loosely explained as being related to the Fermi functions of the leads.
The TEC is zero in the following cases: half filling, where the Fermi function of the leads 
is equal to 0.5, and integer filling, where the Fermi function of the leads is either 0 or 1~\cite{PhysRevB.46.9667}.

For instance, the TEC is zero at $\mu = 1.25$~meV corresponding to 
approximately half filling of the ground state as is shown in \fig{fig03}(b). 
In addition, the TEC is again zero at $\mu = 0.7$ and $1.7$~meV for 
the integer filling of occupation 0 and 1, respectively.

In the presence of a higher temperature of the left lead at $k_B T_L = 0.25$~meV,
a shifting in the TEC is observed for the first-excited state at $\mu = 1.99$~meV.
The current deviation occurs due to an increased thermal smearing at the higher temperature.  In this case, 
both the ground state and the first-excited state participate in the electron transport at $\mu = 1.99$~meV.
Therefore, the current becomes negative, instead of the zero value at a lower termperature.

Furthermore, we notice that the second-excited state of the quantum wire is in resonance with the second subband of the leads.
The wavefunction of the second-excited state of the central system is symmetric in the $y$-direction 
while the wavefunctions of the second subband of the leads are anti-symmetric in the $y$-direction.
The electron transport from a symmetric to an anti-symmetric state or vice versa is not allowed 
due to the geometry sensitive function that describes the coupling between the quantum 
wire and the leads~\cite{Vidar11.113007,Nzar.25.465302}. Therefore, a plateau in the current is formed for the 
second-excited state around $\mu = 3.23$~meV.

\begin{figure}[htbq]
  \includegraphics[width=0.5\textwidth,angle=0,bb=50 95 410 225]{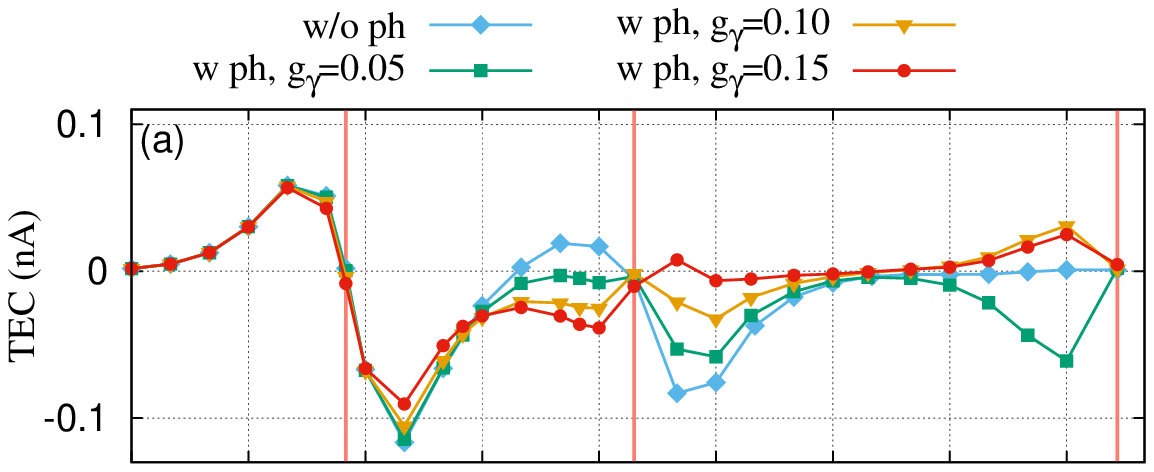}\\
  \includegraphics[width=0.5\textwidth]{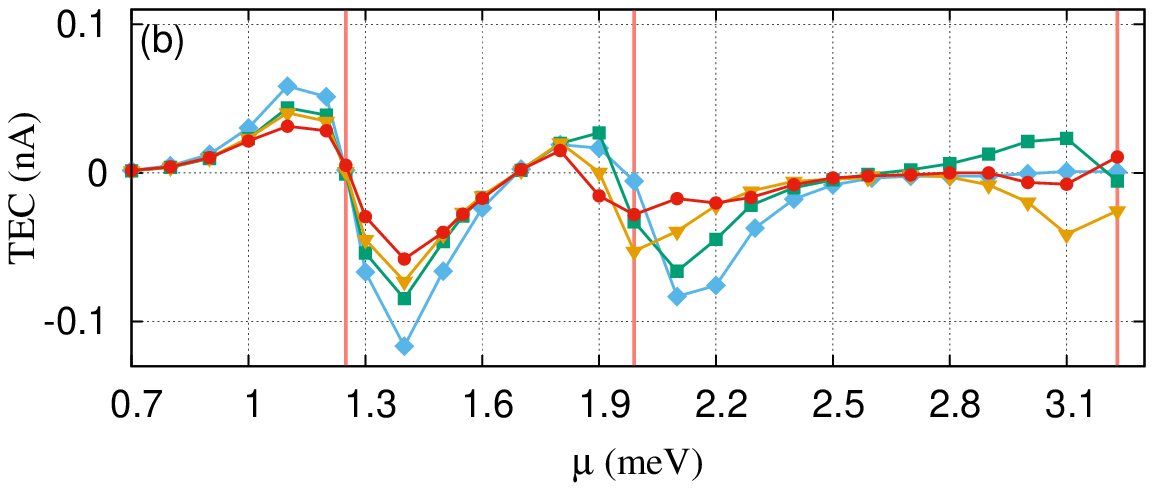}
 \caption{(Color online) TEC as a function of the chemical potential 
                         $\mu = \mu_L = \mu_R$ plotted at time $t = 220$~ps
                         without photon cavity $g_{\gamma} = 0.0$~meV (blue diamonds), 
                         and with photon cavity for the electron-photon coupling strength 
                         $g_{\gamma} = 0.05$ (green squares), $0.10$ (golden triangles), 
                         and $0.15$~meV (red circles) in the case of off-resonance (a) and resonance (b) photon field.
                         The photon energy is $\hbar \omega_{\gamma} = 0.3$~meV and 
                         the photons are linearly polarized in the $x$-direction.
                         The temperature of the left (right) lead is fixed at $T_L = 1.16$~K ($T_R = 0.58$~K) implying 
                         thermal energy $k_B T_L = 0.10$~meV ($k_B T_R = 0.05$~meV), respectively.
                         The magnetic field is $B = 0.1~{\rm T}$, and $\hbar \Omega_0 = 2.0~{\rm meV}$.}
\label{fig04}
\end{figure}

Now, we assume the quantum wire is embedded in a photon cavity with a photon mode of
energy $\hbar\omega_{\gamma} = 0.3$~meV, and the cavity initially contains two photons $N_{\gamma} = 2$. 
The photons are linearly polarized in the $x$-direction.
The photon energy is smaller than the energy spacing between the ground state and the first-excited state on one hand, 
and the first-excited state and the second-excited state on the other hand. This means that the system 
is off-resonant with respect to the photon field. The energy spectrum of the quantum wire embedded in 
the cavity with off-resonant photon field was shown in \fig{fig02}(a). 

The TEC as a function of the chemical potential of the leads is displayed in \fig{fig04}(a) for the system
without a photon cavity $g_{\gamma} = 0.0$~meV (blue diamonds), and with a photon cavity for the electron-photon coupling strength 
$g_{\gamma} = 0.05$ (green squares), $0.10$ (golden triangles), and $0.15$~meV (red circles) in the case of off-resonant photon field
at time $t = 220$~ps.
We observe that the current through the ground state is almost unchanged in the presence of a photon cavity,
but the characteristics of the TEC of the first-excited state and the second-excited state are drastically modified.
The influence of cavity photon field is to invert the TEC from `positive' to `negative' values or vice versa
around the first-excited state at $\mu = 1.99$~meV. The two photon replica of the ground state at $\mu = 1.84$~meV contributes 
to the electron transport with the first-excited state leading to the current flip from `positive' to `negative' values.
In addition, the two photon replica of the first-excited state around $\mu = 2.59$~meV becomes active in the transport.
Therefore, the TEC is inverted from `negative' to `positive' values around $\mu = 2.1$~meV.

\begin{figure}[htbq]
  \includegraphics[width=0.5\textwidth,angle=0,bb=50 95 410 240]{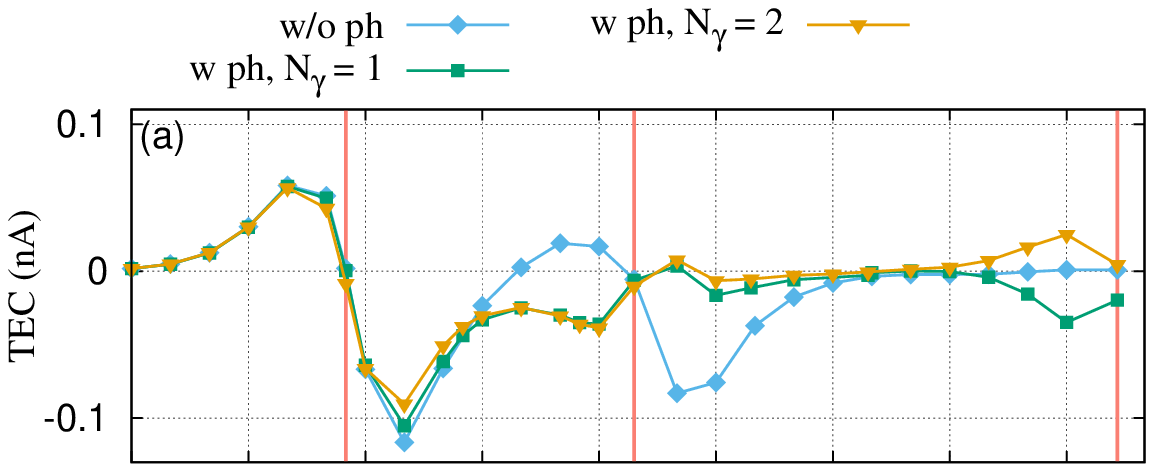}\\
  \includegraphics[width=0.5\textwidth]{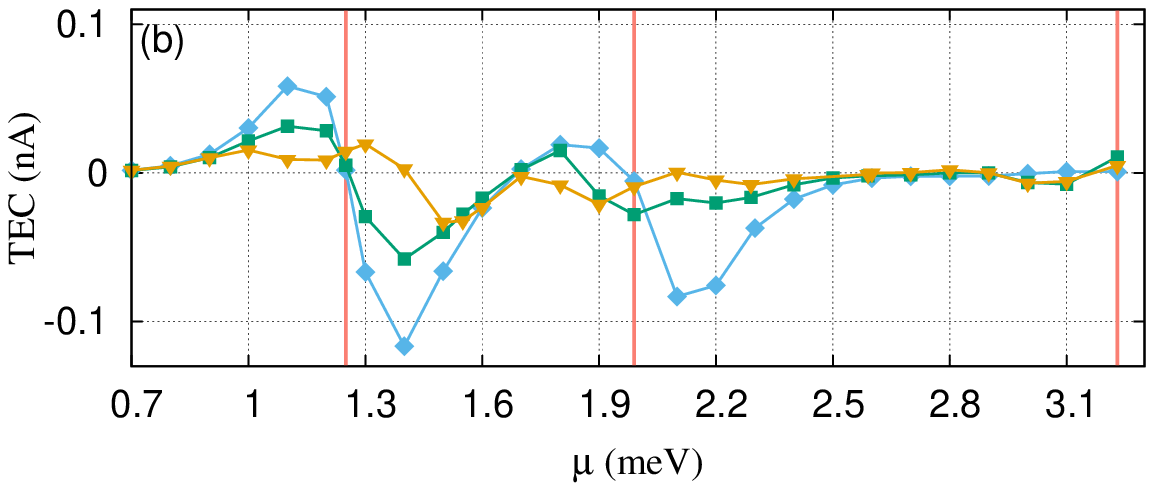}
 \caption{(Color online) TEC as a function of the chemical potential 
                         $\mu = \mu_L = \mu_R$ plotted at time $t = 220$~ps
                         without photon cavity $g_{\gamma} = 0.0$~meV (blue diamonds), 
                         and with photon cavity initially containing one photon $N_{\gamma} = 1$ (green squares)
                         and two photons $N_{\gamma} = 2$ (golden triangles)
                         in the case of off-resonance (a) and resonance (b) photon field.
                         The photon energy is $\hbar \omega_{\gamma} = 0.3$~meV and 
                         the photons are linearly polarized in the $x$-direction.
                         The temperature of the left (right) lead is fixed at $T_L = 1.16$~K ($T_R = 0.58$~K) implying 
                         thermal energy $k_B T_L = 0.10$~meV ($k_B T_R =  0.05$~meV), respectively.
                         The magnetic field is $B = 0.1~{\rm T}$, and $\hbar \Omega_0 = 2.0~{\rm meV}$.}
\label{fig05}
\end{figure}

The electron transport is affected around the second-excited state in the presence of the photon cavity as is shown in \fig{fig04}(a).
This is because the two photon replica of the first-excited state is getting into resonance with the second subband of the leads.
The two photon replica of the first-excited state has an anti-symmetric wavefunction in the $y$-direction and the 
wavefunctions of the second 
subband are anti-symmetric as well. Consequently, the electrons transfer from the second subband of the leads to the 
two photon replica of the first-excited state of the quantum wire. A TEC is thus generated.

The effects of the initial number of photons on the TEC is shown in Fig.\ \ref{fig05}. The off-resonance system
(Fig.\ \ref{fig05}(a)) is rather insensitive to the exact number of photons in the low energy regime.  
This is because both one and two photon replica states are close to each other in the off-resonant photon field.
The thermal smearing leads to participation of both states in the electron transport. Therefore, the selected initial photon number in the cavity
does not influence the TEC.

\subsection{On-resonance photon field}
In this section we assume the photon energy is approximately equal to the energy 
spacing between the ground-state and the first-excited state $\hbar\omega_{\gamma} \cong E_1 - E_0$.
The system under consideration is in resonance with the photon field.
The photon energy is assumed to be $\hbar\omega_{\gamma} = 0.74$~meV and the cavity initially contains one photon, 
$N_{\gamma} = 1$.

The energy spectrum of the quantum wire as a function of the
electron-photon coupling strength $g_{\gamma}$ in the presence of the resonant photon field
is displayed in \fig{fig02}(b), 
where 0ES (golden circles) are zero-electron states 
and 1ES (blue squares) are one-electron states. The horizontal lines (red lines) display the energy of 
possible transport resonances.
In the case of no electron-photon coupling at $g_{\gamma} = 0.0$~meV the three states mentioned in the previous section 
are again found in the selected range of energy. In the presence of the cavity field
the one photon replica of the ground state is observed near the first-excited state at $g_{\gamma} = 0.05$~meV.

Increasing the electron-photon coupling strength to $g_{\gamma} = 0.15$~meV, 
the one photon replica of the ground state is lowered in energy and the first-excited state shifts up.
Therefore, the separation between these two states increases. Similar splitting in 
the energy spectrum can be seen in the higher energy states between $2.5\text{-}3.0$~meV.
The splitting is the Rabi-splitting. 

Figure \ref{fig04}(b) shows TEC versus chemical potential 
at time $t = 220$~ps without a photon cavity $g_{\gamma} = 0.0$~meV (blue diamonds) 
and with a photon cavity for the electron-photon coupling strength 
$g_{\gamma} = 0.05$ (green squares), $0.10$ (golden triangles), and $0.15$~meV (red circles) 
in the case of resonant of the photon field.
The temperature of the left lead is fixed at $T_L = 1.16$~K implying 
thermal energy $k_B T_L = 0.10$~meV and 
the temperature of the right lead is assumed to be $T_R = 0.58$~K with
thermal energy $k_B T_R = 0.05$~meV.

In the resonant photon field, a reduction in the TEC is observed 
with increasing electron-photon coupling strength. 
For $g_{\gamma} = 0.05$~meV, the following three states contribute to the electron transport at $\mu = 1.10$ meV
with `positive' current, and at $\mu = 1.40$ meV with `negative' current: The ground state, the one photon replica 
of the ground state, and the first-excited state. 
Increasing the electron-photon coupling strength to $g_{\gamma} = 0.15$~meV the first-excited state 
shifts up and does not participate in the electron transport. Therefore, the TEC is suppressed.

The TEC decreases at $\mu = 1.90$ and $2.10$ meV around the first-excited state.
The current suppression is due to participation of the one photon replica of 
the first-excited state with the energy $2.65$~meV at $g_{\gamma} = 0.05$~meV. 
But at the higher electron-photon coupling strength 
$g_{\gamma} = 0.15$~meV the one photon replica of the first-excited state is not active 
in the transport. This is because it moves up for high electron-photon coupling strength.
Consequently, the TEC drops down. This reduction in the thermoelectric
current is a direct consequence of the Rabi-splitting of the energy levels of the system.
Earlier, we have established dynamic effects of the Rabi-splitting on the transport current
through the system at a finite bias voltage \cite{Vidar-ACS-Phot}.
The electrons are getting active in the transport around the second-excited state. 
The activation of the electron transport there is due to the symmetry properties of the one photon replica 
of the first-excited state.
We have verified that the TEC flattens out as the temperature of both
leads is increased keeping their temperature difference constant.

Opposite to the off-resonant system, the initial photon number in the resonant system can change 
the TEC substantially, as is seen in Fig.\ \ref{fig05}(b). The reason is the 
larger separation between photon replica states here. Consequently, the place of the active photon 
replicas in the energy spectrum is important 
and the initial photon number influences the TEC.

\section{Conclusions}\label{Sec:Conclusions}
We studied numerically the thermoelectric transport properties
of a short quantum wire interacting with either off-
or on-resonant cavity-photon field.
The quantum wire is assumed to be connected to two electron reservoirs
with different temperatures. The temperature gradient accelerates electrons
from the leads to the quantum wire creating a thermoelectric current.
We showed that a plateau in the current is formed due to symmetry
properties of the energy states of the quantum wire and the leads in
the case of no cavity photon field.
By applying a linearly polarized photon field, the photon replica states
result in an inverted thermoelectric transport in the
off-resonance regime. Moreover, in the on-resonant photon field
the effects of a Rabi-splitting in the energy spectrum appears 
leading to a reduction
in the thermoelectric current. In both regimes, the current plateaus that
were formed in the absence of a photon field are removed due to 
the generation of photon replica states.
Our results point to new opportunities to experimentally 
control thermoelectric transport properties of nanodevices
with a cavity photon field.

\begin{acknowledgments}

Financial support is acknowledged from the Icelandic Research and Instruments Funds, 
and the Research Fund of the University of Iceland. The calculations were carried out on 
the Nordic High Performance Computer Center in Iceland. We acknowledge the Nordic network
NANOCONTROL, project No.: P-13053, and the Ministry of Science and Technology, Taiwan 
through Contract No. MOST 103-2112-M-239-001-MY3.

\end{acknowledgments}

%
%

%
\end{document}